\documentclass{article}
\usepackage[T1]{fontenc}
\usepackage{geometry}
\usepackage{graphics}
\usepackage{setspace}
\makeatletter

\newcommand{\LyX}{L\kern-.1667em\lower.25em\hbox{Y}\kern-.125emX\@}
\newcommand{\noun}[1]{\textsc{#1}}

\makeatother

\begin{document}

{\par\centering {\LARGE Classical and Quantum Oscillators of Sextic and Octic
Anharmonicities }\LARGE \par}

~\\
~\\
~\\

{\par\centering {\large Anirban Pathak and Swapan Mandal }\large \par}

~

{\par\centering {\large Department of Physics}\large \par}

{\par\centering {\large Visva-Bharati}\large \par}

{\par\centering {\large Santiniketan-731235, }\large \par}

{\par\centering {\large INDIA} \par}

~

\begin{abstract}
Classical oscillators of sextic and octic anharmonicities are solved analytically
up to the linear power of \( \lambda  \) \( (Anharmonic\, Constant) \) by
using Taylor series method. These solutions exhibit the presence of secular
terms which are summed up for all orders. The frequency shifts of the oscillators
for small anharmonic constants are obtained. It is found that the calculated
shifts agree nicely with the available results to-date. The solutions for classical
anharmonic oscillators are used to obtain the solutions corresponding to quantum
anharmonic oscillators by imposing fundamental commutation relations between
position and momentum operators. \\
\\
\newpage

\end{abstract}

\section{Introduction:}

\bigskip{}
The basic understanding of Physics is based on some physical models. The simple
harmonic oscillator (SHO) is perhaps the most useful one among them. A particle
subject to a restoring force proportional to its displacement gives rise to
the model of a SHO. The Hamiltonian corresponding to a classical SHO of unit
mass and unit frequency is given by 
\begin{equation}
\label{1a}
H\, =\, \frac{p^{2}}{2}\, +\, \frac{x^{2}}{2}\: 
\end{equation}
and the Hamilton's equations are 
\begin{equation}
\label{1b}
\begin{array}{lcl}
\dot{p} & = & -\frac{\partial H}{\partial x}=-x\\
\dot{x} & = & \frac{\partial H}{\partial p}=p
\end{array}.
\end{equation}
The equation of motion of the SHO is 
\begin{equation}
\label{1c}
\ddot{x}\, +\, x\, =\, 0
\end{equation}
 where, equation (\ref{1b}) is used. Now the solution of the equation (\ref{1c})
is found to be
\begin{equation}
\label{1d}
x_{0}(t)=x(0)cost+\dot{x}(0)sint.
\end{equation}
 The parameters \( x(0) \) and \( \dot{x}(0) \) are the initial position and
momentum of the oscillator. Thus the position of the oscillator at a later time
\( t \) is completely known in terms of the initial position and momentum.
Nevertheless, the momentum is also known from the Hamilton's equations (\ref{1b}).
Hence, the oscillator problem is completely solved. 

The equation of motion of a quantum SHO may simply be obtained by imposing \( x \)
and \( \dot{x} \) as noncommuting operators. To avoid confusion, we represent
the operators as capital letters instead of lowercase letters. Thus the noncommuting
position and momentum operators obey the following relation 
\begin{equation}
\label{1e}
[X(t),\dot{X}(t)]\, =\, i
\end{equation}
 where \( \hbar =1 \). The solution of a quantum SHO in Heisenberg picture
is the operator equivalent of the equation (\ref{1d}). Now for real physical
problems, the anharmonicity and/or damping are to be incorporated in the model
Hamiltonian and hence in the equation of motion. However, the present letter
takes care the anharmonicity only. The general form of the Hamiltonian of an
anharmonic oscillator (having unit mass and unit frequency) of even order of
anharmonicity is given by 
\begin{equation}
\label{1g}
H\, =\, \frac{p^{2}}{2}+\frac{x^{2}}{2}+\frac{\lambda }{2m}x^{2m}
\end{equation}
 where \( m\, (\geq 2) \) is an integer and \( \lambda  \) is the anharmonic
constant. Depending upon the problem of physical interests, different type of
anharmonic oscillators will appear. For \( m=2, \) the corresponding oscillator
is called quartic anharmonic oscillator. Similarly, \( m=3 \) and \( m=4 \)
give rise to the sextic and octic anharmonic oscillators respectively. Of course,
the anharmonic constants \( \lambda  \) are different for different type of
anharmonic oscillators. The equation of motion corresponding to the Hamiltonian
(\ref{1g}) is 
\begin{equation}
\label{1h}
\ddot{x}+x+\lambda x^{2m-1}=0.
\end{equation}
 where the equation (\ref{1b}) is used.

\section{Quartic Oscillator: Brief Review. }

The problem of classical and quantum anharmonic oscillators have already been
studied by several authors. To begin with, we recall the problem of a classical
quartic oscillator (Duffing oscillator). The equation (\ref{1h}), for \( m=2, \)
does not have exact solution. However, large number of approximate methods are
available for the purpose of getting analytical solutions to the Duffing oscillator
problem. These include perturbation technique {[}\ref{Nayfeh}{]}, variation
of parameters {[}\ref{Ross}{]} and Taylor series approaches {[}\ref{Marganeau}{]}.
The ordinary perturbation technique, a pedestrian approach, leads to the unwanted
secular terms {[}\ref{Nayfeh}{]}. The removal of secular terms from the solution
is a serious problem. There are some methods which are successfully used to
summed up the secular terms for all orders. These include ``tucking in technique''
{[}\ref{Bellmann}{]}, multiscale perturbation theory and renormalization technique{[}\ref{Nayfeh}{]}. 

The search for the solutions of anharmonic oscillators enriched different brunches
of Physics and Mathematics, for example, it is successfully used to enrich the
subject of large order perturbation theory {[}\ref{Bender 1}-\ref{Lowdin}{]},
divergent expansion of quantum mechanics {[}\ref{Simon}-\ref{Graffi}{]}, Laplace
transformation representation of energy eigenvalues {[}\ref{Ivanov}{]} and
computational physics {[}\ref{Weniger}-\ref{Fernandez}{]}. 

The solutions of quantum quartic anharmonic oscillators may be obtained in two
different pictures. The Schroedinger picture deals with the time development
of the wave function. On the other hand, the time development of operators are
obtained in the Heisenberg picture. Basically, the Schroedinger picture is used
to solve the quantum oscillators as eigenvalue problems. In these problems,
energy eigenvalues are expressed as the sum of different orders of anharmonic
constant \( \lambda  \). The energy eigenvalues are found to diverge for large
anharmonic constant. In case of small \( \lambda  \), however, the eigenvalues
for different orders are summed up and the convergence of these sums are ensured
by the Borel summability {[}\ref{Graffi},\ref{Graffi 1}{]} and/ or Stieljet
conditions {[}\ref{Loeffel}{]}. 

The c-number approach to the quantum oscillator problem is found very successful
but the situation is different with the operator approach. Actually, the noncommuting
nature of the operators (\ref{1e}) pose a serious difficulty for the purpose
of getting approximate solutions to the quantum anharmonic oscillator problems.
Only few methods are available to-date. Aks {[}\ref{aks0}{]}and Aks and Caharat
{[}\ref{Aks}{]} first obtained an operator solution of a quantum quartic anharmonic
oscillator by using the method of Bogoliubov and Krylov. But the recent interest
in the problem started with the work of Bender and Bettencourt {[}\ref{Bender}-\ref{cmb1}{]}
who obtained the solution of a quantum quartic anharmonic oscillator by using
multiscale perturbation theory (MSPT). Later on Mandal{[}\ref{Mandal}{]} proposed
a Taylor series approach, Egusquiza and Basagoiti{[}\ref{egusquiza}{]} developed
renormalization technique, Kahn and Zarmi{[}\ref{kahn}{]} used a near identity
transform method to solve the quantum quartic anharmonic oscillator problem.
The Taylor series approach is found to be more successful compared to the other
approaches for the purpose of getting second or higher order solution of a quantum
quartic anharmonic oscillator {[}\ref{Anirban}{]}. 

In addition to the quartic oscillator, people have studied the higher anharmonic
oscillators also {[}\ref{Weniger}, \ref{Laksman}-\ref{Dutt}{]}. These studies
are made to obtain the eigenvalues as function of anharmonicity. More recently
Pathak {[}\ref{pathak}{]} used a normal ordering technique to obtain a zeroth
order multiscale perturbation theoretic solution of \( m-th \) anharmonic oscillator
and Fernandez{[}\ref{fernandez}{]} developed an eigen value approach to construct
first order correction to the frequency operator for \( m-th \) anharmonic
oscillator. However, a first order solution of higher anharmonic oscillators
in the realm of Heisenberg approach are yet to come. Keeping that in mind, the
present letter will take care the analytical solutions of classical and quantum
oscillators of sextic and octic anharmonicities. These oscillators have potential
applications in the studies of nonlinear mechanics, molecular physics, quantum
optics and in the field theory. This is the first time we give analytical solution
of sextic and octic oscillators under the operator formalism. The solutions
are used to obtain the frequency shifts of sextic and octic oscillators. The
computed shifts are compared and found to have exact coincidence with the frequency
shifts calculated by using a first order perturbation theory and frequency operators
obtained by Pathak {[}\ref{pathak}{]}.

\section{Sextic Oscillator. }

For \( m=3, \) the Hamiltonian (\ref{1g}) and the equation of motion (\ref{1h})
correspond to the case of a sextic anharmonic oscillator. The solution for such
an oscillator is obtained as the sum of different orders of anharmonicities.
The corresponding solution is 
\begin{equation}
\label{2}
x(t)\, =x_{0}(t)\, +\lambda x_{1}(t)\, +\, \ldots 
\end{equation}
 where, \( x_{0}(t) \) and \( x_{1}(t) \) are zeroth and first order solutions
respectively. The zeroth order solution (\ref{1d}) is simply obtained for \( \lambda =0 \)
in equation (\ref{1h}). The purpose of the present section is to find the first
order solution \( x_{1}(t). \) The lower case \( x(t) \) and \( \dot{x}(t) \)
are to be treated as time developments of position and momentum of the classical
sextic anharmonic oscillator. The equivalent operator representations are to
be made by using upper case letters \( X(t) \) and \( \dot{X}(t) \). Thus
the solution of the classical sextic anharmonic oscillator is given by the following
Taylor series
\begin{equation}
\label{3}
x(t)=x(0)\, +t\dot{x}(0)\, +\frac{t^{2}}{2!}\ddot{x}(0)+\ldots 
\end{equation}
Here we assume that \( t \) is sufficiently small and a series (\ref{3}) expansion
is possible. Now  we express higher order time derivatives of \( x(t) \) at
\( t=0 \) as
\begin{equation}
\label{4}
\begin{array}{lcl}
x^{iii}(0) & = & -\dot{x}(0)-5\lambda x^{4}(0)\dot{x}(0)\\
x^{iv}(0) & = & x(0)+6\lambda x^{5}(0)-20\lambda x^{3}(0)\dot{x}^{2}(0)\\
x^{v}(0) & = & \dot{x}(0)+70\lambda x^{4}(0)\dot{x}(0)-60\lambda \dot{x}^{3}(0)x^{2}(0)\\
... & ... & ...\\
\ldots  & \ldots  & \ldots 
\end{array}
\end{equation}
where, we have neglected \( \lambda ^{2} \) and higher order terms. We substitute
the relations (\ref{4}) in equation (\ref{3}) to collect the coefficients
of \( x^{5}(0),\, x^{4}(0)\dot{x}(0),\, x^{3}(0)\dot{x}^{2}(0),\, x^{2}(0)\dot{x}^{3}(0),\, x(0)\dot{x}^{4}(0)\,  \)
and \( \dot{x}^{5}(0) \) (without the constant multiplication factor \( \lambda  \)).
Corresponding coefficients are 
\begin{equation}
\label{5}
\begin{array}{lcl}
C_{0} & = & -(1-6+71-1276+27741-656546+\ldots )\\
C_{1} & = & -(5-70+1275-27740+656545-\ldots )\\
C_{2} & = & -(20-460+10680-258920+6399340-\ldots )\\
C_{3} & = & -(60-1860+49320-12577720+31664580+\ldots )\\
C_{4} & = & -(120-4320+120240-3116640+78912360-\ldots )\\
and &  & \\
C_{5} & = & -(120-4320+120240-3116640+78912360-\ldots )
\end{array}
\end{equation}
 where the coefficient of \( x^{5-i}(0)\dot{x}^{i}(0) \) is denoted by \( C_{i} \).
The contributions directly from Taylor expansion are not taken into consideration.
The \( r-th \) term of \( C_{0} \) is given by 
\begin{equation}
\label{6}
\begin{array}{lcl}
t^{/}_{r} & = & \left( -1\right) ^{r}\times \frac{1}{384}\left( 25^{r}+15\times 9^{r}+240r-16\right) .\\
 &  & 
\end{array}
\end{equation}
Now we can write the \( r-th \) term of the coefficient of \( x^{5}(0) \)
in \( x_{1}(t) \) as 
\begin{equation}
\label{7}
t_{r}=t^{/}_{r}\times \frac{t^{2r}}{(2r)!},
\end{equation}
 where the factor \( \frac{t^{2r}}{(2r)!} \) comes from the Taylor expansion
part. The net coefficient of \( x^{5}(0) \) in \( x_{1}(t) \) is obtained
as 
\begin{equation}
\label{8}
K_{0}=\sum t_{r}=\frac{1}{384}(cos5t+15\, cos3t-16\, cost-120\, tsint).
\end{equation}
The parameter \( K_{0} \) exhibits fifth and third harmonic generations along
with the secular term proportional to \( t\, sint. \) The secular terms are
responsible for the divergent nature of the solution. This is an uncomfortable
situation and has to be taken care properly in order to get a well behaved solution
for all orders. Now, the remaining coefficients are obtained by using the same
procedure as adopted for the evaluation of \( K_{0}. \) The corresponding coefficients
are 

\begin{equation}
\label{9}
\begin{array}{lcl}
K_{1} & = & \frac{1}{384}(5sin5t+45sin3t-280sint+120tcost)\\
K_{2} & = & \frac{2}{384}(-5cos5t-15cos3t+20cost-120tsint)\\
K_{3} & = & \frac{2}{384}(-5sin5t+15sin3t-140sint+120tcost)\\
K_{4} & = & \frac{1}{384}(5cos5t-45cos3t+40cost-120tsint)\\
and &  & \\
K_{5} & = & \frac{1}{384}(sin5t-15sin3t-80sint+120tcost)
\end{array}
\end{equation}
Hence, the first order solution is given by
\begin{equation}
\label{10}
x_{1}(t)=\sum ^{5}_{i=0}K_{i}x^{5-i}(0)\dot{x}^{i}(0).
\end{equation}
Thus the total solution (\ref{2}) is simply 
\begin{equation}
\label{changed1}
x(t)=x(0)cost+\dot{x}(0)sint+\lambda \sum ^{5}_{i=0}K_{i}x^{5-i}(0)\dot{x}^{i}(0).
\end{equation}

Now to have a check of the solution (\ref{10}), we consider a special case
\( \dot{x}(0)=0 \) and \( x(0)=a \). Hence the total solution reduces to an
extraordinarily simple form as 
\begin{equation}
\label{11}
x(t)=acost+\frac{\lambda a^{5}}{384}(cos5t+15\, cos3t-16\, cost-120\, tsint).
\end{equation}
It is clear that the presence of secular term proportional to \( tsint \) makes
the life difficult as \( t \) grows. We confine ourselves in the weak coupling
regime and we assume \( C \) is a constant such that
\begin{equation}
\label{13a}
\begin{array}{lcl}
sin\left( C\lambda t\right)  & = & C\lambda t\\
cos\left( C\lambda t\right)  & = & 1
\end{array}.
\end{equation}
 Now, the equation (\ref{11}) is rearranged as 
\begin{equation}
\label{12}
x(t)=acos[t(1+\frac{5}{16}\lambda a^{4})]+\lambda a^{5}(-\frac{1}{24}cost+\frac{5}{128}cos3t+\frac{1}{384}cos5t)
\end{equation}
where the equation (\ref{13a}) is used. Thus the secular term present in the
first order solution (\ref{11}) is summed up for all orders and the frequency
shift of the oscillator is obtained. The corresponding frequency shift of the
sextic anharmonic oscillator is \( \frac{5}{16}\lambda a^{4} \). The shifted
frequency of the oscillator is also viewed as the renormalization of frequency
by the anharmonic interaction. Interestingly, the frequency shift obtained in
the solution (\ref{12}) exactly coincides with that obtained by Dutt and Lakshmanan
{[}\ref{Dutt}{]}. Now to have an idea of how good the solutions (\ref{11})
and (\ref{12}) are we compare them with the exact numerical solution obtained
by Mathematica (figure 1). From figure 1, we observe that the first order frequency
renormalized solution (\ref{12}) coincides exactly with the exact numerical
solution obtained by using Mathematica (for \( a=2 \) and \( \lambda =0.01 \)).
However, the solution (\ref{11}) diverges with the increase of time (figure
1). The divergent nature of the solution manifests the presence of secular term.

\vspace{0.375cm}
{\par\centering \resizebox*{6in}{4in}{\rotatebox{270}{\includegraphics{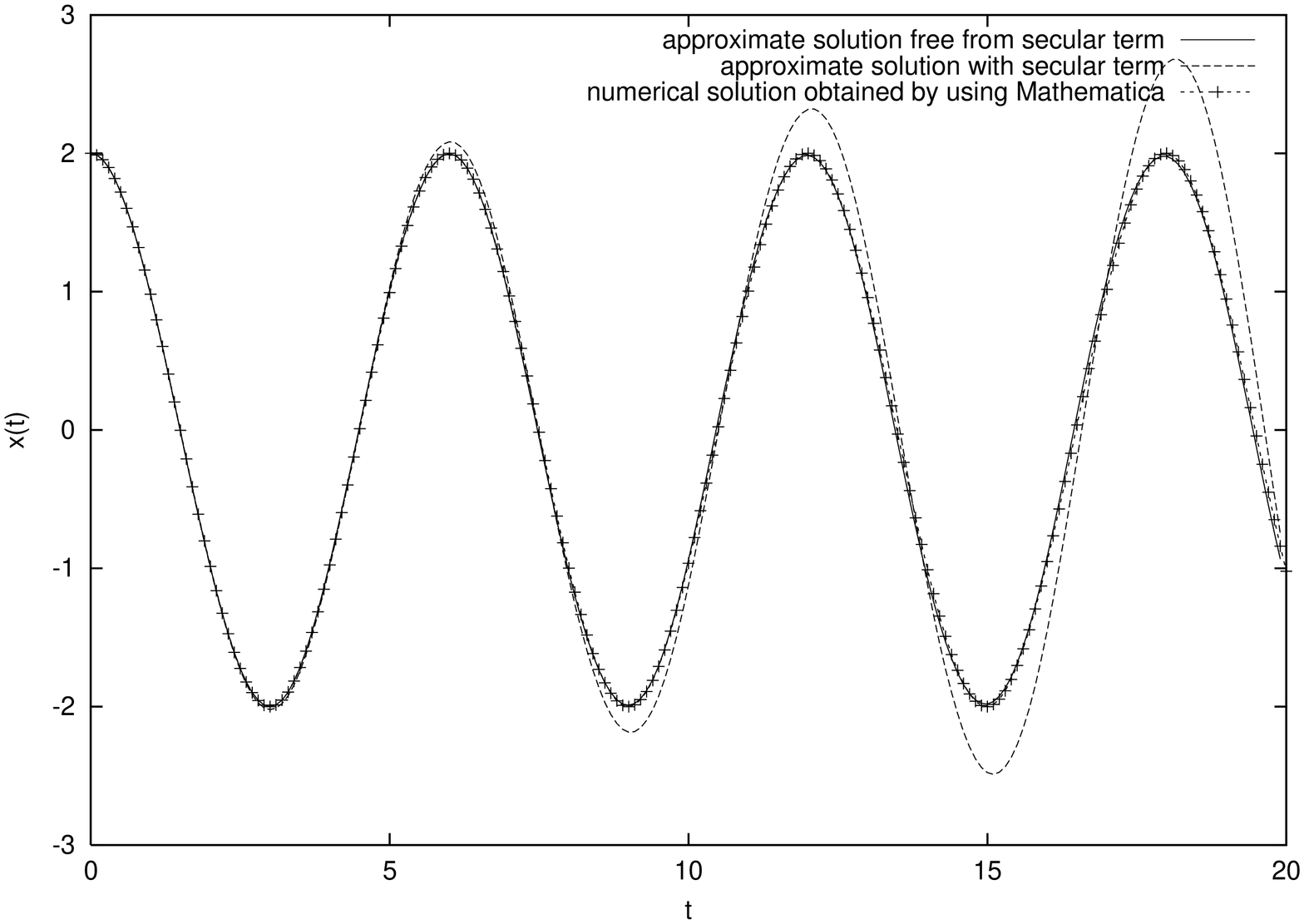}}} \par}
\vspace{0.375cm}

{\par\centering figure1\par}

Now we replace \( x(t) \) and \( \dot{x}(t) \) with the noncommuting operators
\( X(t) \) and \( \dot{X}(t) \) respectively. Finally, it is straightforward
to generalize the classical solution (\ref{2}) into a quantum solution of the
anharmonic oscillator by the imposition of commutation relation (\ref{1e}).
The corresponding solution of the quantum sextic oscillator is given by

\begin{equation}
\label{14}
\begin{array}{lcl}
X(t) & = & X(0)cost+\dot{X}(0)sint+\frac{\lambda }{384}X^{5}(0)\times (cos5t+15\, cos3t-16\, cost-120\, tsint)\\
 & + & \frac{\lambda }{1920}[X^{4}(0)\dot{X}(0)+X^{3}(0)\dot{X}(0)X(0)+X^{2}(0)\dot{X}(0)X^{2}(0)+X(0)\dot{X}(0)X^{3}(0)+\dot{X}(0)X^{4}(0)]\\
 & \times  & (5sin5t+45sin3t-280sint+120tcost)\\
 & + & \frac{\lambda }{1920}[X^{3}(0)\dot{X}^{2}(0)+X^{2}(0)\dot{X}^{2}(0)X(0)+X^{2}(0)\dot{X}(0)X(0)\dot{X}(0)\\
 & + & X(0)\dot{X}^{2}(0)X^{2}(0)+X(0)\dot{X}(0)X(0)\dot{X}(0)X(0)+X(0)\dot{X}(0)X^{2}(0)\dot{X}(0)\\
 & + & \dot{X}^{2}(0)X^{3}(0)+\dot{X}(0)X(0)\dot{X}(0)X^{2}(0)+\dot{X}(0)X^{2}(0)\dot{X}(0)X(0)+\dot{X}(0)X^{3}(0)\dot{X}(0)]\\
 & \times  & (-5cos5t-15cos3t+20cost-120tsint)\\
 & + & \frac{\lambda }{1920}[\dot{X}^{3}(0)X^{2}(0)+\dot{X}^{2}(0)X^{2}(0)\dot{X}(0)+\dot{X}^{2}(0)X(0)\dot{X}(0)X(0)+\dot{X}(0)X^{2}(0)\dot{X}^{2}(0)\\
 & + & \dot{X}(0)X(0)\dot{X}(0)X(0)\dot{X}(0)+\dot{X}(0)X(0)\dot{X}^{2}(0)X(0)+X^{2}(0)\dot{X}^{3}(0)\\
 & + & X(0)\dot{X}(0)X(0)\dot{X}^{2}(0)+X(0)\dot{X}^{2}(0)X(0)\dot{X}(0)+X(0)\dot{X}^{3}(0)X(0)]\\
 & \times  & (-5sin5t+15sin3t-140sint+120tcost)\\
 & + & \frac{\lambda }{1920}[X(0)\dot{X}^{4}(0)+\dot{X}(0)X(0)\dot{X}^{3}(0)+\dot{X}^{2}(0)X(0)\dot{X}^{2}(0)+\dot{X}^{3}(0)X(0)\dot{X}(0)+\dot{X}^{4}(0)X(0)]\\
 &  & \\
 & \times  & (5cos5t-45cos3t+40cost-120tsint)\\
 & + & \frac{\lambda \dot{X}^{5}(0)}{384}(sin5t-15sin3t-80sint+120tcost)\\
 &  & 
\end{array}
\end{equation}
where \( X(0) \) and \( \dot{X}(0) \) are the initial position and momentum
operators. The solution (\ref{14}) may be written in the symmetrical form as
\begin{equation}
\label{15}
\begin{array}{lcl}
X(t) & = & X(0)cost+\dot{X}(0)sint\\
 & + & \frac{2\lambda }{768}X^{5}(0)(cos5t+15\, cos3t-16\, cost-120\, tsint)\\
 & + & \frac{\lambda }{768}[X^{4}(0)\dot{X}(0)+\dot{X}(0)X^{4}(0)]\\
 & \times  & (5sin5t+45sin3t-280sint+120tcost)\\
 & + & \frac{2\lambda }{768}[X^{3}(0)\dot{X}^{2}(0)+\dot{X}^{2}(0)X^{3}(0)+3X(0)]\\
 & \times  & (-5cos5t-15cos3t+20cost-120tsint)\\
 & + & \frac{2\lambda }{768}[\dot{X}^{3}(0)X^{2}(0)+X^{2}(0)\dot{X}^{3}(0)+3\dot{X}(0)]\\
 & \times  & (-5sin5t+15sin3t-140sint+120tcost)\\
 & + & \frac{\lambda }{768}[X(0)\dot{X}^{4}(0)+\dot{X}^{4}(0)X(0)]\\
 & \times  & (5cos5t-45cos3t+40cost-120tsint)\\
 & + & \frac{2\lambda \dot{X}^{5}(0)}{768}(sin5t-15sin3t-80sint+120tcost)\\
 &  & 
\end{array}
\end{equation}
 The equation (\ref{15}) is our desired solution for a quantum sextic anharmonic
oscillator. Now we rearrange our derived solution (\ref{15}) with the help
of (\ref{13a} ) as
\begin{equation}
\label{16}
\begin{array}{lcl}
X(t) & = & \frac{1}{2}\left\{ X(0)cos\left( t+\frac{5\lambda t}{4}\left\{ H^{2}_{0}+\frac{1}{4}\right\} \right) +cos\left( t+\frac{5\lambda t}{4}\left\{ H^{2}_{0}+\frac{1}{4}\right\} \right) X(0)\right. \\
 & + & \dot{X}(0)sin\left( t+\frac{5\lambda t}{4}\left\{ H^{2}_{0}+\frac{1}{4}\right\} \right) +sin\left( t+\frac{5\lambda t}{4}\left\{ H^{2}_{0}+\frac{1}{4}\right\} \right) \dot{X}(0)\\
 & + & \frac{2\lambda X^{5}(0)}{384}(cos5t+15\, cos3t-16\, cost)\\
 & + & \frac{\lambda }{384}[X^{4}(0)\dot{X}(0)+\dot{X}(0)X^{4}(0)](5sin5t+45sin3t-280sint)\\
 & + & \frac{2\lambda }{384}[X^{3}(0)\dot{X}^{2}(0)+\dot{X}^{2}(0)X^{3}(0)+3X(0)](-5cos5t-15cos3t+20cost)\\
 & + & \frac{2\lambda }{384}[\dot{X}^{3}(0)X^{2}(0)+X^{2}(0)\dot{X}^{3}(0)+3\dot{X}(0)](-5sin5t+15sin3t-140sint)\\
 & + & \frac{\lambda }{384}[X(0)\dot{X}^{4}(0)+\dot{X}^{4}(0)X(0)](5cos5t-45cos3t+40cost)\\
 & + & \left. \frac{2\lambda \dot{X}^{5}(0)}{384}(sin5t-15sin3t-80sint)\right\} \\
 &  & 
\end{array}
\end{equation}
 where \( H_{0}=\frac{\dot{X}^{2}(0)}{2}+\frac{X^{2}(0)}{2} \) is the unperturbed
Hamiltonian. Now the first order frequency operator, \( \Omega =1+\frac{5\lambda }{4}\left\{ H^{2}_{0}+\frac{1}{4}\right\}  \)
coincides exactly with the MSPT results of Pathak {[}\ref{pathak}{]}. The matrix
element \( <n-1|X(t)|n> \) and hence the frequency shift of the oscillator
may also be calculated with the help of following relations 
\begin{equation}
\label{17}
\begin{array}{lcl}
H_{0}|n> & = & (n+\frac{1}{2})|n>\\
X(0)|n> & = & \frac{1}{\sqrt{2}}\left[ \sqrt{n}|n-1>+\sqrt{n+1}|n+1>\right] \, for\, n\neq 0\\
\dot{X}(0)|n> & = & \frac{-i}{\sqrt{2}}\left[ \sqrt{n}|n-1>-\sqrt{n+1}|n+1>\right] \, for\, n\neq 0.
\end{array}
\end{equation}
From the field theoretic point of view, \( n \) may be regarded as the number
of photons present in the quantized radiation field. Now, the dipole moment
matrix element in terms of the number eigenket \( |n> \) is
\begin{equation}
\label{18}
\begin{array}{lcl}
<n-1|X(t)|n> & = & \frac{cos(\frac{5\lambda t}{4}n)}{2}\left\{ <n-1|X(0)|n>cos\left( t+\frac{5\lambda t}{4}\left\{ n^{2}+\frac{1}{2}\right\} \right) \right. \\
 & + & cos\left( t+\frac{5\lambda t}{4}\left\{ n^{2}+\frac{1}{2}\right\} \right) <n-1|X(0)|n>\\
 & + & <n-1|\dot{X}(0)|n>sin\left( t+\frac{5\lambda t}{4}\left\{ n^{2}+\frac{1}{2}\right\} \right) \\
 & + & sin\left( t+\frac{5\lambda t}{4}\left\{ n^{2}+\frac{1}{2}\right\} \right) <n-1|\dot{X}(0)|n>\\
 & + & <n-1|\left[ \frac{2\lambda X^{5}(0)}{384}(cos5t+15\, cos3t-16\, cost)\right. \\
 & + & \frac{\lambda }{384}[X^{4}(0)\dot{X}(0)+\dot{X}(0)X^{4}(0)](5sin5t+45sin3t-280sint)\\
 & + & \frac{2\lambda }{384}[X^{3}(0)\dot{X}^{2}(0)+\dot{X}^{2}(0)X^{3}(0)+3X(0)](-5cos5t-15cos3t+20cost)\\
 & + & \frac{2\lambda }{384}[\dot{X}^{3}(0)X^{2}(0)+X^{2}(0)\dot{X}^{3}(0)+3\dot{X}(0)](-5sin5t+15sin3t-140sint)\\
 & + & \frac{\lambda }{384}[X(0)\dot{X}^{4}(0)+\dot{X}^{4}(0)X(0)](5cos5t-45cos3t+40cost)\\
 & + & \left. \left. \frac{2\lambda \dot{X}^{5}(0)}{384}(sin5t-15sin3t-80sint)\right] |n>\right\} .\\
 &  & 
\end{array}
\end{equation}
 Hence, we obtain the frequency shift of the oscillator as \( \frac{5\lambda }{4}(n^{2}+\frac{1}{2}). \)
It is interesting to note that the frequency shift of the oscillator is possible
even when there are no photons (i.e vacuum field) present in the radiation field.
This phenomena is a direct outcome of pure quantum electrodynamic effect and
has no classical analog. The frequency shift of the oscillator for a vacuum
field is actually a second order effect and is reported elsewhere {[}\ref{Anirban}{]}.
The solution (\ref{16}) is of its first kind and does not have any direct check.
However, the frequency shift in (\ref{18}) may be compared with the frequency
shift calculated by using first order perturbation technique {[}\ref{Powel}{]}.
To have a check, we calculate the energy of a sextic oscillator in \( n-th \)
energy state which is given by
\begin{equation}
\label{15a}
E_{n}=(n+\frac{1}{2})+\frac{5\lambda }{48}(4n^{3}+6n^{2}+8n+3).
\end{equation}
 The energy difference between two consecutive levels is 
\begin{equation}
\label{15b}
\Delta E=E_{n}-E_{n-1}=1+\frac{5\lambda }{4}(n^{2}+\frac{1}{2}).
\end{equation}
 Thus our calculated frequency shift and hence the solution agrees nicely. Finally
we normalize the equation (\ref{16})with the help of the equation (\ref{18})
as
\begin{equation}
\label{19}
\begin{array}{lcl}
X(t) & = & \frac{1}{2cos(\frac{5\lambda t}{4}n)}\left\{ X(0)cos\left( t+\frac{5\lambda t}{4}\left\{ H^{2}_{0}+\frac{1}{4}\right\} \right) +cos\left( t+\frac{5\lambda t}{4}\left\{ H^{2}_{0}+\frac{1}{4}\right\} \right) X(0)\right. \\
 & + & \dot{X}(0)sin\left( t+\frac{5\lambda t}{4}\left\{ H^{2}_{0}+\frac{1}{4}\right\} \right) +sin\left( t+\frac{5\lambda t}{4}\left\{ H^{2}_{0}+\frac{1}{4}\right\} \right) \dot{X}(0)\\
 & + & \frac{2\lambda X^{5}(0)}{384}(cos5t+15\, cos3t-16\, cost)\\
 & + & \frac{\lambda }{384}[X^{4}(0)\dot{X}(0)+\dot{X}(0)X^{4}(0)](5sin5t+45sin3t-280sint)\\
 & + & \frac{2\lambda }{384}[X^{3}(0)\dot{X}^{2}(0)+\dot{X}^{2}(0)X^{3}(0)+3X(0)](-5cos5t-15cos3t+20cost)\\
 & + & \frac{2\lambda }{384}[\dot{X}^{3}(0)X^{2}(0)+X^{2}(0)\dot{X}^{3}(0)+3\dot{X}(0)](-5sin5t+15sin3t-140sint)\\
 & + & \frac{\lambda }{384}[X(0)\dot{X}^{4}(0)+\dot{X}^{4}(0)X(0)](5cos5t-45cos3t+40cost)\\
 & + & \left. \frac{2\lambda \dot{X}^{5}(0)}{384}(sin5t-15sin3t-80sint)\right\} .\\
 &  & 
\end{array}
\end{equation}
 The equation (\ref{19}) is the desired solution for quantum sextic anharmonic
oscillator where the secular terms for all orders are summed up.

\section{Octic Oscillator. }

Depending upon the nature of nonlinearity, the higher anharmonic oscillators
will come into picture. For example, octic anharmonic oscillator may be obtained
if we put \( m=4 \) in the equations (\ref{1g}) and (\ref{1h}) respectively.
Using the similar procedure (as adopted for sextic oscillator), the solution
(up to the linear power of \( \lambda  \)) for an octic oscillator follows
immediately as 
\begin{equation}
\label{oc2}
\begin{array}{lcl}
x(t) & = & x(0)cost+\dot{x}(0)sint\\
 & + & \frac{\lambda x^{7}(0)}{3072}(cos7t+14cos5t+126cos3t-141cost-840tsint)\\
 & + & \frac{\lambda \dot{x}(0)x^{6}(0)}{3072}(7sin7t+70sin5t+378sint-2373sint+840tcost)\\
 & + & \frac{\lambda \dot{x}^{2}(0)x^{5}(0)}{3072}(-21cos7t-126cos5t-126cos3t+273cost-2520tsint)\\
 & + & \frac{\lambda \dot{x}^{3}(0)x^{4}(0)}{3072}(-35sin7t-70sin5t+630sin3t-3815sint+2520tcost)\\
 & + & \frac{\lambda \dot{x}^{4}(0)x^{3}(0)}{3072}(35cos7t-70cos5t-630cos3t+665cost-2520tsint)\\
 & + & \frac{\lambda \dot{x}^{5}(0)x^{2}(0)}{3072}(21sin7t-126sin5t+126sin3t-2415sint+2520tcost)\\
 & + & \frac{\lambda \dot{x}^{6}(0)x(0)}{3072}(-7cos7t+70cos5t-378cos3t+315cost-840tsint)\\
 & + & \frac{\lambda \dot{x}^{7}(0)}{3072}(-sin7t+14sin5t-126sin3t-525sint+840tcost)\\
 & 
\end{array}
\end{equation}
 To check the validity of our solution, we consider a special case \( x(0)=b \)
and \( \dot{x}(0)=0. \) After a little algebra we have 
\begin{equation}
\label{oc3}
x(t)=bcos(1+\frac{35\lambda }{128}b^{6})t+\frac{\lambda b^{7}}{3072}(-141cost+126cos3t+14cos5t+cos7t).
\end{equation}
 Thus the frequency shift of the classical octic oscillator is proportional
to the sixth power of amplitude as long as the first order solution is concerned.
The solution (\ref{oc3}) as well as the frequency shift have exact coincidence
with the solution obtained by the procedure introduced by Bradbury and Brintzenhoff
{[}\ref{Bradley}{]}. Now we generalize the solution (\ref{oc2}) to have the
corresponding quantum solution in the symmetrical form as
\begin{equation}
\label{oc4}
\begin{array}{lcl}
X(t) & = & X(0)cost+\dot{X}(0)sint\\
 & + & \frac{\lambda X^{7}(0)}{3072}(cos7t+14cos5t+126cos3t-141cost-840tsint)\\
 & + & \frac{\lambda }{6144}[\dot{X}(0)X^{6}(0)+X^{6}(0)\dot{X}(0)](7sin7t+70sin5t+378sint-2373sint+840tcost)\\
 & + & \frac{\lambda }{6144}[\dot{X}^{5}(0)X^{2}(0)+X^{2}(0)\dot{X}^{5}(0)+10X^{3}(0)\}\\
 & \times  & (-21cos7t-126cos5t-126cos3t+273cost-2520tsint)\\
 & + & \frac{\lambda }{6144}[\dot{X}^{4}(0)X^{3}(0)+X^{3}(0)\dot{X}^{4}(0)+9(X^{2}(0)\dot{X}(0)+\dot{X}(0)X^{2}(0))]\\
 & \times  & (-35sin7t-70sin5t+630sin3t-3815sint+2520tcost)\\
 & + & \frac{\lambda }{6144}[\dot{X}^{3}(0)X^{4}(0)+X^{4}(0)\dot{X}^{3}(0)+9(\dot{X}^{2}(0)X(0)+X(0)\dot{X}^{2}(0))]\\
 & \times  & (35cos7t-70cos5t-630cos3t+665cost-2520tsint)\\
 & + & \frac{\lambda }{6144}[\dot{X}^{5}(0)X^{2}(0)+X^{2}(0)\dot{X}^{5}(0)+10\dot{X}^{3}(0)]\\
 & \times  & (21sin7t-126sin5t+126sin3t-2415sint+2520tcost)\\
 & + & \frac{\lambda }{3072}[\dot{X}^{6}(0)X(0)+X(0)\dot{X}^{6}(0)](-7cos7t+70cos5t-378cos3t+315cost-840tsint)\\
 & + & \frac{\lambda \dot{X}^{7}(0)}{3072}(-sin7t+14sin5t-126sin3t-525sint+840tcost)
\end{array}
\end{equation}
 The problem of secular terms are taken care by the same procedure as it is
done in the case of a sextic anharmonic oscillator. The final solution appears
as 
\begin{equation}
\label{oc5}
\begin{array}{lcl}
X(t) & = & \frac{1}{2cos[\frac{35\lambda }{64}(6n^{2}+3)t]}\left\{ X(0)cos\Omega t+cos\Omega tX(0)+\dot{X}(0)sin\Omega t+sin\Omega t\dot{X}(0)\right. \\
 & + & \frac{2\lambda X^{7}(0)}{3072}(cos7t+14cos5t+126cos3t-141cost)\\
 & + & \frac{\lambda }{3072}[\dot{X}(0)X^{6}(0)+X^{6}(0)\dot{X}(0)](7sin7t+70sin5t+378sint-2373sint)\\
 & + & \frac{\lambda }{3072}[\dot{X}^{5}(0)X^{2}(0)+X^{2}(0)\dot{X}^{5}(0)+10X^{3}(0)]\\
 & \times  & (-21cos7t-126cos5t-126cos3t+273cost)\\
 & + & \frac{\lambda }{3072}[\dot{X}^{4}(0)X^{3}(0)+X^{3}(0)\dot{X}^{4}(0)+9(X^{2}(0)\dot{X}(0)+\dot{X}(0)X^{2}(0))]\\
 & \times  & (-35sin7t-70sin5t+630sin3t-3815sint)\\
 & + & \frac{\lambda }{3072}[\dot{X}^{3}(0)X^{4}(0)+X^{4}(0)\dot{X}^{3}(0)+9(\dot{X}^{2}(0)X(0)+X(0)\dot{X}^{2}(0))]\\
 & \times  & (35cos7t-70cos5t-630cos3t+665cost)\\
 & + & \frac{\lambda }{3072}[\dot{X}^{5}(0)X^{2}(0)+X^{2}(0)\dot{X}^{5}(0)+10\dot{X}^{3}(0)]\\
 & \times  & (21sin7t-126sin5t+126sin3t-2415sint)\\
 & + & \frac{\lambda }{3072}[\dot{X}^{6}(0)X(0)+X(0)\dot{X}^{6}(0)](-7cos7t+70cos5t-378cos3t+315cost)\\
 & + & \left. \frac{2\lambda \dot{X}^{7}(0)}{3072}(-sin7t+14sin5t-126sin3t-525sint)\right\} \\
 & 
\end{array}
\end{equation}
  where \( \Omega =1+\frac{35\lambda }{64}(4H_{0}^{3}+5H_{0}) \), is the first
order frequency operator which is in exact accordance with the frequency operator
obtained by Pathak{[}

{]}. Now, the matrix element \( <n-1|X(t)|n> \) is calculated to obtain the
shifted frequency as \( \omega ^{/}=1+\frac{35\lambda }{16}(n^{3}+2n) \). Interestingly,
the frequency shift \( \frac{35\lambda }{16}(n^{3}+2n) \) depends on \( n^{3}. \)
It is clear that the vacuum field has no role to play in the frequency shift
of the quantum octic oscillator. This is the first time we give the solution
of an octic oscillator (\ref{oc5}) and therefore, it is not possible to have
a direct verification of the solution. However, to have a possible verification
of the solution, we calculate the frequency shift by using Rayleigh-Schroedinger
perturbation theory. A first order calculation under this theory shows that
the energy of the octic oscillator in \( n-th \) state is given by 
\begin{equation}
\label{oc6}
E_{n}=(n+\frac{1}{2})+\frac{35\lambda }{64}(\frac{3}{2}+4n+5n^{2}+2n^{3}+n^{4}).
\end{equation}
 The energy difference between two consecutive levels is \( \Delta E\, =\, E_{n}-E_{n-1}=1+\frac{35\lambda }{16}(n^{3}+2n). \)
Thus, the frequency shift exactly coincides with the frequency shift obtained
by us.

\section{Remarks and Conclusions: }

We obtain analytical solutions (first order) for sextic (\ref{2}) and octic
(\ref{oc2}) oscillators in classical pictures by using Taylor series method.
The solutions exhibit the presence of usual secular terms. The secular terms
for all orders are tucked in to obtain the frequency shifts of those oscillators.
It is found that the calculated shifts agree satisfactorily with the shifts
obtained by other methods. 

The position and momentum of classical oscillators are replaced by their corresponding
operators and the commutation relation is imposed. Hence, the solutions for
quantum sextic and octic oscillators in the realm of Heisenberg picture is obtained.
Interestingly, the vacuum field has no role to play in the frequency shift of
the quantum oscillators of quartic and octic anharmonicities. However, the vacuum
field causes the frequency shift of a quantum sextic oscillator. These apparent
discrepancies are easily understood in terms of field theoretic point of view
where the vacuum field effect is considered as a second order one. In fact,
the frequency shift due to vacuum field is obtained in a second order solution
(i.e solution contains \( \lambda ^{2}) \) of a quantum quartic oscillator
{[}\ref{Anirban}{]}. 

To the best of our knowledge, there only few possible methods available for
the purpose of getting analytical solutions of a quantum quartic anharmonic
oscillators {[}\ref{aks0}-\ref{Anirban}{]}. The method devised by one of us
{[}\ref{Mandal}{]} is already tested its usefulness for the purpose of getting
an analytical second order solution of a quantum quartic anharmonic oscillator
{[}\ref{Anirban}{]}. However, there is no publication till date which deal
with the solution of higher anharmonic oscillators. Therefore, the solutions
for higher anharmonic oscillators of the present letter will serve a lot of
academic interest. In addition to these the solutions will be useful in nonlinear
mechanics, molecular physics and in the field theory. In particular, the solutions
for quantum anharmonic oscillators are useful to investigate the quantum statistical
properties of radiation field. For example, the squeezing effects are reported
for a third order nonlinear non-absorbing medium by using the solution of the
quantum quartic oscillator {[}\ref{Anirban},\ref{mandalb}{]}. 

The present solutions along with the solutions given by us {[}\ref{Mandal},\ref{pathak}{]}
seem to be useful for the purpose of getting a general solution of \( m-th \)
anharmonicities. Such solutions will be taken care in future publications. 

~\\
\\
\textbf{Acknowledgments}: \emph{One of the authors (AP) is thankful to the CSIR
for the award of a Senior Research Fellowship. The work is supported by the
Department of Science and Technology (SR/SY/P-10/92) and the Council of Scientific
and Industrial Research (03/798/96-EMR-II)}.\\
\\
\\
\textbf{~~~References}

\begin{enumerate}
\item \label{Nayfeh}A. H. Nayfeh, Introduction to Perturbation Techniques (Wiley,
New York, 1981) p.112. 
\item \label{Ross} S. L. Ross , \emph{}Differential Equation 3rd Eds. (John Wiley,
New York, 1984) p.707.
\item \label{Marganeau}H. Margenau and G. M. Murphy, The Mathematics of Physics and
Chemistry \emph{(}D.Van Nostrand Company, Princeton, 1956\emph{)} p.483. 
\item \label{Bellmann}R. Bellman, Methods of Nonlinear Analysis Vol.1 \emph{(}Academic
Press, New York, 1970\emph{)} p.198 . 
\item \label{Bender 1}C. M. Bender and T. T. Wu, Phys.Rev., 184 (1969) 1231. 
\item \label{Bender 2}C. M. Bender and T. T. Wu, Phys.Rev.Lett\emph{.}, 27 (1971)
461.
\item \label{Loeffel} J. J. Loeffel, A. Martin, B. Simon and A. S. Wightman, Phys.Lett
\emph{}30B (1969) 656. 
\item \label{Bender 3}C. M. Bender and T. T. Wu, Phys.Rev D\emph{.}, 7 (1973) 1620.
\item \label{Arteca}G. A. Arteca, F. M. Fernandez and E. A. Castro, Large order Perturbation
Theory and Summation Methods in Quantum Mechanics \emph{}(Berlin: Springer-Verlag,
1990). 
\item \label{Lowdin}P. O. Lowdin (Ed.), Int.J.Quanum.Chem. \emph{}(1982)\emph{,}
(Proceedings of the Sanibel Workshop on Perturbation Theory at Large Order,
Sanibel Conference, Florida, 1981)
\item \label{Simon}B. Simon, \emph{}Ann. Phys\emph{.}, A 31 (1970) 76.
\item \label{Simon2}B. Simon, Int. J. Quantum Chemistry, \emph{}21 (1982) 3.
\item \label{Graffi}S. Graffi, V. Greechi and B. Simon, Phys. Lett., 32B (1970) 631.
\item \label{Ivanov}I. A. Ivanov, J Phys. A, 31 (1998) 5697.
\item \label{Weniger}E. J. Weniger, Ann.Phys., 246 (1996) 133. 
\item \label{Weniger 1}E. J. Weniger, J. Cizek and F. Vinette, \emph{}J. Math. Phys\emph{.,}
39 (1993) 571.
\item \label{Vinette}F. Vinette and J. Cizek, J. Math. Phys.\emph{,} 32 (1991) 3392.
\item \label{Fernandez}F. Fernandez and J. Cizek, \emph{}Phys. Lett., \emph{}166A
(1992) 173.
\item \label{Graffi 1}S. Graffi, V. Greechi and G. Turchetti, \emph{}IL Nuovo Cimento,
\emph{}4B (1971) 313. 
\item \label{aks0} S O Aks, Fortsch. der Phys. \textbf{15} (1967) 661.
\item \label{Aks}S O Aks and Caharat R A, IL Nuovo Cimento \textbf{64} (1969) 798.
\item \label{Bender}C. M. Bender and L. M. A. Bettencourt, Phys. Rev. Lett\emph{.,}
77 (1996) 4114. 
\item \label{cmb1} C. M. Bender and L. M. A. Bettencourt, Phys. Rev. D\emph{,} 54
(1996) 7710. 
\item \label{Mandal}S. Mandal, J. Phys. \emph{}A, 31 (1998) L501.
\item \label{egusquiza}Egusquiza IL and Valle Basagoitti M A, Phys. Rev. A \textbf{57}
(1998)1586.
\item \label{kahn}P B Kahn and Zarmi Y, J. Math. Phys. \textbf{40} (1998) 4658.
\item \label{Anirban} A. Pathak and S. Mandal, Phys. Lett.A \textbf{286} (2001) 261
\item \label{Laksman}M.Lakshmanan and J.Prabhakaran, \noun{}\emph{\noun{Lett.Nuovo.Cimento
7}} (1973) 689. 
\item \label{Biswas}S. N. Biswas, K. Datta, R. P. Saxena, P. K. Srivastava and V.
S. Verma, J.Math.Phys, 14 (1973) 1190. 
\item \label{Dutt}R. Dutt and M. Lakshmanan, J.Math.Phys 17 (1976) 482. 
\item \label{pathak}A Pathak, J. Phys A \textbf{33} (2000) 5607.
\item \label{fernandez}F M Fernandez, J. Phys A \textbf{34} (2001) 4851.
\item \label{Bradley}T. C. Bradbury and A. Brintzenhoff, J.Math.Phys 12 (1971) 1269
. 
\item \label{Powel} J. L. Powel and B. Crasemann, Quantum Mechanics (Addision-Wesley\textbf{,}
Massachusetts, 1961)
\item \label{mandalb}S. Mandal, J.Phys.B \textbf{33} (2000) 1029. 
\end{enumerate}
\end{document}